# Measuring the Competitive Pressure of Academic Journals and the Competitive Intensity within Subjects

Ma Zheng[1], Yu Zhenglu[2], Pan Yuntao[3], Wu Yishan[4]

*[1] mazheng@istic.ac.cn*
Institute of Scientific and Technical Information of China (ISTIC), 15 Fuxing Rd. 100038 Beijing (China)
Nanjing University, School of information Management, 22 Hankou Rd. 210093 Nanjing (China)

*[2] luluyu@istic.ac.cn*
Institute of Scientific and Technical Information of China (ISTIC), 15 Fuxing Rd. 100038 Beijing (China)

*[3] panyt@istic.ac.cn*
Institute of Scientific and Technical Information of China (ISTIC), 15 Fuxing Rd. 100038 Beijing (China)

*[4] wuyishan@istic.ac.cn*
Institute of Scientific and Technical Information of China (ISTIC), 15 Fuxing Rd. 100038 Beijing (China)

**Abstract**
A journal's impact and similarity with rivals is closely related to its competitive intensity. A subject area can be considered as an ecological system of journals, and can then be measured using the competitive intensity concept from plant systems. Based on Journal Citation Reports data from 1997, 2000, 2005, 2010, and 2013, we calculated the mutual citation, cosine similarity, and competitive relationship matrices for mycology journals. We derived the mutual citation network for mycology according to Journal Citation Reports data from 2013. We calculated each journal's competitive pressure, and the competitive intensity for the subject. We found that competitive pressures are very variable among journals. Differences between a journal's absolute and relative influence are related to the competitive pressure. A more powerful journal has lower competitive pressure. New journals have more competitive pressure. If there are no other influences, the competition intensity of a subject will continue to increase. Furthermore, we found that if a subject has more journals, its competitive intensity decreases.

**Introduction**

Scientific and technical (S&T) journals have an important role in science and knowledge dissemination. Journals that are focussed on the same subject are at competition with each other. We must build a favourable competitive environment to realize the optimal allocation of limited resources. At the same time, the "survival of the fittest" mechanism boosts the development of S&T journals.
To build a sustainable environment and competition mechanism, we must analyse and measure the present environment of S&T journals, especially in terms of competition. Many researchers have investigated the competitive environment of S&T journals.

*Reaching a consensus on the relationship between the journal environment and competition*
Scholars began to study the competitive relationship of journals in the 1920s. Competition is mainly related to the resources of subeditors, editors, and authors. Studies found that competitive power is related to a journals' impact factor (IF) (Campanario 1996). Zhu (1999) discussed the relationship between an S&T journal's quality and competitive spirit. A few years later, scholars proposed that competition is a basic attribute of science and noted the differences between different journals' abilities to secure resources. Powerful journals typically attract more attention, which results in a Matthew effect on the journal's development. Scholars have attempted to measure competition between journals using quantitative indexes (Manfred and Scharnhorst 2001). Researchers have generally accepted that S&T journals develop within a competitive environment. They have explored definitions of the competition between S&T journals (Cai 2003), how to increase a journal's core competitive strength (Chen 2005), and how take advantage of market competition (Gao 2004). Recently, Leydesdorff et al. focused on competition between highly cited journals dependent on the proportions of most-frequently cited publications in the European Union, China, and the United States, which are represented differently because they use different databases. (Leydesdorff. 2014)

*Determining the competitive relationship between journals using quantitative methods*

Leydesdorff noted that Pearson correlations could be used as similarity measures for citation patterns based on bi-connected graphs (Leydesdorff 2004). He then used principal component analysis and factor analysis to design indicators for the position of the cited journals in the dimensions of the database (Leydesdorff 2006). Yang analysed the relationship between a journal's value chain and competitive edge using value chain theory (Yang 2006). As a whole, these ideas and methods for quantitatively measuring a journal's competitive relationship have not been generally accepted, and are not fully developed.

*Applying research ideas from ecological competition*

Recently, ideas related to competition and competitive intensity in ecology have been applied to research related to S&T journals. Scholars such as Tao have attempted to consider the survival and development of S&T journals from an ecological perspective (Tao 2007). Xinyan researched the concentration ratio of an S&T journal's market share and its competition. She also analysed the index model of competitive intensity in ecology, and applied it to measure a journal's competitive intensity (CI). This was a meaningful exploration, but did not result in a proper index for measuring a journal's distance in terms of the ecological system of S&T journals (Xinyan 2008).

The competitive environment of S&T journals has been extensively analysed. Progress has been made in terms of the quantitative analysis. Although the CI concept from ecology is useful, we do not know how to define and measure the "distance" between journals. The institute of Scientific and Technical Information of China has measured journal similarity using the mutual citation matrix and cosine similarity method since 2011 (ISTIC 2011). This provides a measurement of the distance between journals.

In this study, we considered a journal's absolute impact value and similarity as parameters based on the *Journal Citation Reports*. We measured the competitive pressures of mycology journals and the CI for the entire subject using scientometrics and the CI.

**Methodology**

In this study, we used the concept of CI from the field of ecological research to define the "competitive pressure" among S&T journals. The following design scheme illustrates how we calculate the relevant values.

*Main factors that influence the competitive relationship between S&T journals*

In a relatively closed ecological environment, the CI mainly depends on the differences between plant diameters and the distance between plants. In this closed environment, the competitive relationships between plants can indicate the strength of the overall competition within the ecological environment.

If we consider journals that focus on one subject, we are investigating a relatively closed ecological environment. Then, all the individual journals can be viewed as separate plants. As shown in Figure 1, the respective "diameters" ($D_i$ and $D_j$) of journals $i$ and $j$, and the "distance" ($L_{ij}$) between them are the major factors of the competitive relationship.

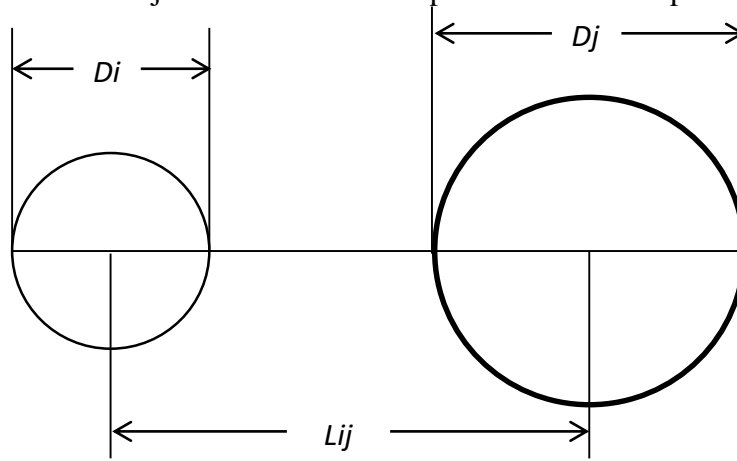

**Figure 1. Main factors influencing the competitive relationship between S&T journals**

*The number of total citations can be used as an alternative indicator to reflect the influence of the journal*

The absolute influence of the journal can be seen as the plant thickness (diameter). Typically, a thicker plant is more capable of competing for resources and fighting rivals. Similarly, more influential journals are generally stronger terms of their access to excellent manuscripts, funding, and attention. Journals with weaker influences are under more pressure from competitors.

The absolute influence of journals can be quantified using three main indicators: total citations (TC), IF, and the number of published papers.

Among these indicators, the IF is more likely to fluctuate. The number of papers is more vulnerable to subjective factors and can sometimes change dramatically. For example, a change to the journal's publishing cycle from bimonthly to monthly will lead to a sudden increase in the number of papers, and an accordingly sharp drop in the IF (because of a doubled denominator). Compared with the IF and paper number, the total citation indicator is relatively more stable and objective. It visually reflects the influence of journals, is less effected by other factors, and has a distinct advantage in terms of long term monitoring.

Additionally, the IF depends on the average number of citations of paper in a journal, so the total citation is equal to the IF multiplied by the number of papers. From this point of view, the total citation is monotonic in the mathematical sense.

Considering the above discussion, the total citation can be used as an alternative indicator of the influence of a journal. Therefore, in this study, we use the total citation as the diameter ($D_i$) of journal $i$. That is,

$$D_i = TC_i, \qquad (1)$$

where $TC_i$ is the total citation of journal $i$.

*The similarity of two journals can be compared using the "distance" between them*

It is widely accepted within the ecological community that competition is most intense when the same species live in the same environment (Clements 1905). The similarity between two journals is also an important factor in their competitive relationship. In other words, a greater similarity between two journals leads to more intense competition. The similarity between two journals can be compared using the "distance" between them ($L_{ij}$).

Zheng et al. calculated a citation matrix for a sample of Chinese journals, which is classified into 61 subjects. They calculated the similarities for each journal in a specific subject area, and then constructed the similarity matrix for the journals (Ma 2012). We used the same definition, and calculated the distance between periodicals using

$$L_{ij} = \frac{1}{S_{ij}} - 1, \qquad (2)$$

where $S_{ij}$ is the cosine similarity indicator between $i$ and $j$. $S_{ij}$ is in the range of [0,1], and $l_{ij}$ is in the range of [0,∞]. A $S_{ij}$ value that is closer to 1 means that journals $i$ and $j$ are more similar. Accordingly, the distance $L_{ij}$ is closer to zero. Conversely, if $S_{ij}$ is closer to zero, $i$ and $j$ are less similar and the distance $L_{ij}$ is closer to infinity.

*Calculating the competition pressure between S&T journals*

We used Hegyi's quantitative measurement for plant competition in ecology (Hegyi 1974). Suppose that there are $n$ journals for a subject, the target journal is called $i$ and is set as the "basic journal", and the other is called $j$ and considered a "rival journal". Then, $CR_{ij}$ is the competitive pressure on journal $i$ from rival $j$. It is calculated using

$$CR_{ij} = \frac{D_j}{D_i \cdot L_{ij}}. \qquad (3)$$

We can assume that the competitive pressure on $i$ from $j$ is inversely proportional to the absolute influence of $i$, is directly proportional to the absolute influence of the rival, and is inversely proportional to the distance between the journals. This assumption is consistent with an intuitive understanding of the competitive relationship.

Combining Equations (1), (2), and (3), we get

$$CR_{ij} = \frac{TC_j}{TC_i \cdot \left(\frac{1}{S_{ij}} - 1\right)}, \qquad (4)$$

where $TC_i$ and $TC_j$ represent the TC for $i$ and $j$, and $S_{ij}$ is the cosine similarity between periodicals.

$CR_{ij}$ and $CR_{ji}$ represent the competitive relationship between $i$ and $j$. The cosine similarity $S_{ij}$ measures the angular distance between a journal and its rival, so $S_{ij}$ and $S_{ji}$ are equal. However, $CR_{ij}$ and $CR_{ji}$ are not equal if $TC_i$ is not equal to $TC_j$. Equation (4) implies that $C_{ij}$ and $C_{ji}$ have a mutually reciprocal relationship.

We can conclude from the definition that the basic journal is under less competitive pressure if it has a higher total citation value than its competitor, and vice versa. The more similar the journals are, the greater the competitive pressure. A journal does not compete with itself, so $CR_{ii}$ is zero.

*Calculating the competitive pressure on basic journal i*

Suppose that, within its discipline, basic journal $i$ has $n-1$ rival journals. Then, $CI_i$ is the total competitive pressure on journal $i$ from all of its rivals,

$$CI_i = \sum_{n}^{j=1} CR_{ij}. \qquad (5)$$

*Overall competitive strength for a specific subject*

The number of competing journals depends on the subject classification. To compare disciplines, we define the overall competitive strength as CIS. It is the average competitive pressure for all journals, i.e.,

$$CIS = \frac{1}{n} \sum_{n}^{i=1} CI_i. \qquad (6)$$

**Analysis and Results**

We calculated the mutual citation, similarity, competitive relationship, and competitive pressure matrices for the journals, and the CI for mycology using Journal Citation Report (*JCR*) data from 1997, 2000, 2005, 2010, and 2003.

*The inter-citation matrices for the target subject, and the similarity and competitive relationships*

We used journals focussed on mycology to demonstrate how to calculate and analyse inter-citations within the target subject, and the similarities and competitive relationships between journals.

There are 23 journals indexed in the *JCR 2013* for mycology (*n=23*). The inter-citation matrix (*C*) was constructed by calculating the inter-citations of each pair of journals. We used the cosine similarity method to transform the inter-citation matrix to the similarity matrix, *R*. The cosine similarity is calculated using

$$\text{Cosine}(x, y) = \frac{\sum_{i=1}^{n} x_i y_i}{\sqrt{\sum_{i=1}^{n} x_i^2} \sqrt{\sum_{i=1}^{n} y_i^2}}. \tag{7}$$

We transformed *R* into a net document and used Pajek to produce Figure 2, which shows the mutual citation network for mycology according to *JCR 2013*. Each node represents a journal, and a node's area represents the journal's TC. The location of the journal and the thickness of the link represent its similarity with its rivals.

From another perspective, we considered the whole subject area as an ecological space. Then, the 23 journals are independent plants. Figure 2 can be regarded as an ecological system with 23 plants, as viewed from above. The differences between the plant diameters and distances between plants determine the CI and the state of the journals.

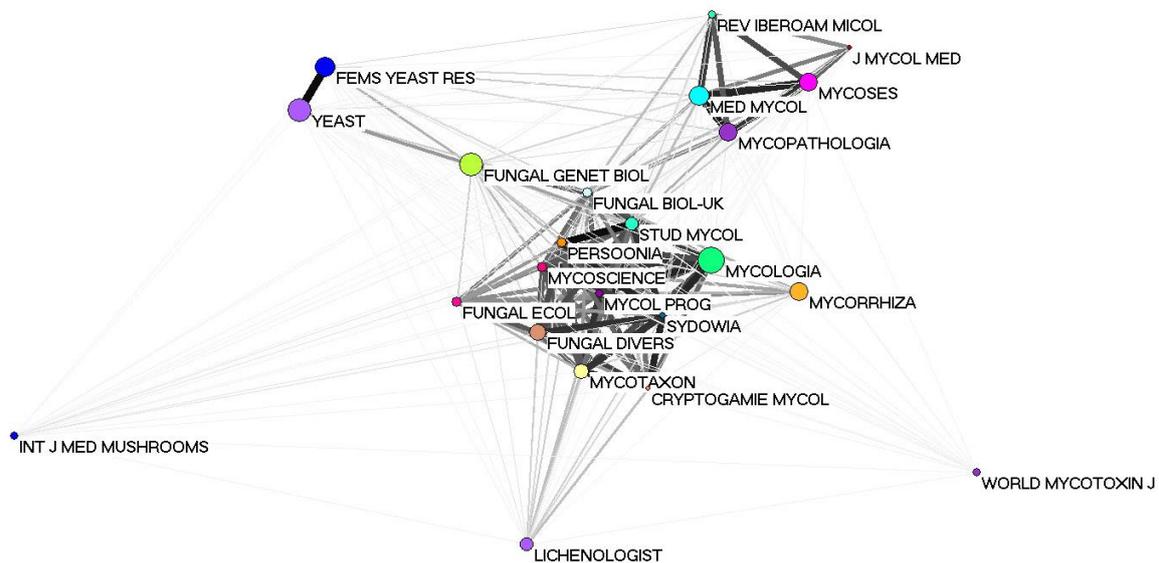

**Figure 2. Mutual citation network of journal focussed on mycology, according to *JCR 2013***

We applied Equation (4) to construct the competitive pressure matrix (*CR*) for the 23 journals, by considering each journal's TC and the cosine similarities between each journal pair.

*Competitive pressure for a journal (CI)*

Equation (5) shows that the CI of a journal is a combination of the competitive pressure from all of each rivals. We measured the competitive pressure of the all journals using competitive relationship matrices for mycology at five time points.

**Table 1. Competitive intensity (CI) for mycology journals**

| Title | 1997 | 2000 | 2005 | 2010 | 2013 |
|---|---|---|---|---|---|
| CRYPTOGAMIE MYCOL | 79.15 | 278.326 | 37.227 | 90.551 | 140.329 |
| EXP MYCOL | 13.81 | | | | |
| FEMS YEAST RES | | | | 17.673 | 32.585 |
| FUNGAL BIOL-UK | | | | 81.125 | 48.575 |
| FUNGAL DIVERS | | | 28.170 | 8.875 | 14.402 |
| FUNGAL ECOL | | | | 16.954 | 23.032 |
| FUNGAL GENET BIOL | 4.394 | 14.820 | 2.985 | 1.929 | 3.222 |
| INT J MED MUSHROOMS | | | | 0.341 | 2.175 |
| J MED VET MYCOL | 13.572 | | | | |
| J MYCOL MED | 42.521 | 18.324 | 31.853 | 17.819 | 41.412 |
| LICHENOLOGIST | | | 3.753 | 3.057 | 3.249 |
| MED MYCOL | | 28.391 | 5.748 | 7.315 | 18.067 |
| MIKOL FITOPATOL | 3.280 | 1.854 | 2.389 | | |
| MYCOL PROG | | | | 189.149 | 98.921 |
| MYCOL RES | 3.751 | 6.649 | 11.217 | 11.919 | |
| MYCOLOGIA | 4.663 | 7.341 | 12.558 | 5.09 | 6.046 |
| MYCOPATHOLOGIA | 11.130 | 4.616 | 5.069 | 6.109 | 17.724 |
| MYCORRHIZA | 4.993 | 8.529 | 4.174 | 2.036 | 2.292 |
| MYCOSCIENCE | | | | 30.886 | 53.764 |
| MYCOSES | 10.392 | 3.991 | 3.422 | 12.211 | 18.333 |
| MYCOTAXON | 16.890 | 20.216 | 18.220 | 15.182 | 16.865 |
| PERSOONIA | 94.223 | 84.520 | 408.198 | | 92.237 |
| REV IBEROAM MICOL | | | | 31.666 | 35.185 |
| STUD MYCOL | 139.528 | 69.935 | 51.901 | 31.591 | 36.342 |
| SYDOWIA | | | 116.148 | 298.986 | 230.812 |
| WORLD MYCOTOXIN J | | | | | 0.095 |
| YEAST | 0.031 | 0.022 | 0.318 | 5.028 | 15.638 |

Table 1 shows that there were large differences in the competitive pressures of the rival journals. The maximum was 408.198 and the minimum was 0.022. In *JCR 2013,* two journals had competitive pressures over 100, 15 were between 10 and 100, and six were under 10.

**Table 2. Competitive intensity (CI) compared with impact factor (IF) and total citations (TC), for mycology journals in 2013**

| Title | CI 2013 | rank | IF 2013 | rank | TC 2013 | rank |
|---|---|---|---|---|---|---|
| CRYPTOGAMIE MYCOL | 140.329 | 2 | 1.153 | 18 | 254 | 22 |
| FEMS YEAST RES | 32.585 | 10 | 2.436 | 7 | 2935 | 5 |
| FUNGAL BIOL-UK | 48.575 | 6 | 2.139 | 10 | 790 | 14 |
| FUNGAL DIVERS | 14.402 | 17 | 6.938 | 2 | 2120 | 9 |
| FUNGAL ECOL | 23.032 | 11 | 2.992 | 5 | 701 | 15 |
| FUNGAL GENET BIOL | 3.222 | 20 | 3.262 | 4 | 4298 | 2 |
| INT J MED MUSHROOMS | 2.175 | 22 | 1.123 | 19 | 554 | 19 |
| J MYCOL MED | 41.412 | 7 | 0.4 | 22 | 247 | 23 |
| LICHENOLOGIST | 3.249 | 19 | 1.613 | 14 | 1285 | 12 |
| MED MYCOL | 18.067 | 13 | 2.261 | 9 | 3132 | 4 |
| MYCOL PROG | 98.921 | 3 | 1.543 | 16 | 623 | 18 |
| MYCOLOGIA | 6.046 | 18 | 2.128 | 11 | 5754 | 1 |

| | | | | | |
|---|---|---|---|---|---|
| MYCOPATHOLOGIA | 17.724 | 14 | 1.545 | 15 | 2913 | 6 |
| MYCORRHIZA | 2.292 | 21 | 2.985 | 6 | 2650 | 7 |
| MYCOSCIENCE | 53.764 | 5 | 1.288 | 17 | 926 | 13 |
| MYCOSES | 18.333 | 12 | 1.805 | 12 | 2451 | 8 |
| MYCOTAXON | 16.865 | 15 | 0.643 | 21 | 1959 | 10 |
| PERSOONIA | 92.237 | 4 | 4.225 | 3 | 669 | 16 |
| REV IBEROAM MICOL | 35.185 | 9 | 0.971 | 20 | 649 | 17 |
| STUD MYCOL | 36.342 | 8 | 9.296 | 1 | 1461 | 11 |
| SYDOWIA | 230.812 | 1 | 0.213 | 23 | 355 | 21 |
| WORLD MYCOTOXIN J | 0.095 | 23 | 2.38 | 8 | 454 | 20 |
| YEAST | 15.638 | 16 | 1.742 | 13 | 4268 | 3 |

Table 2 shows the competitive intensities compared with the IF and TC, for mycology journals in 2013. The rankings based on the IF and TC is different from the CI rankings. Some journals are ranked in the top 10 in terms of TC and IF but have low CIs, and some are ranked in the bottom five in terms of TC and IF but have higher CI*s*. Therefore a more powerful journal has lower competitive pressure. We have only listed the results based on the 2013 data, but they were similar for 1997, 2000, 2005, and 2010. The difference between a journals' absolute and relative influence is related to its competitive pressure.

There are certainly some exceptions. Journals that are extremely similar have a significant influence on the competitive pressure. For example, some journals have TCs that are greater than one thousand and are very similar to other journals with the same mass influence, so they also have high competitive pressures. However, some journals are focused on narrow fields and have distinctive characteristics, and therefore do not have much competition because there are not many similar journals, although their TC may be high.

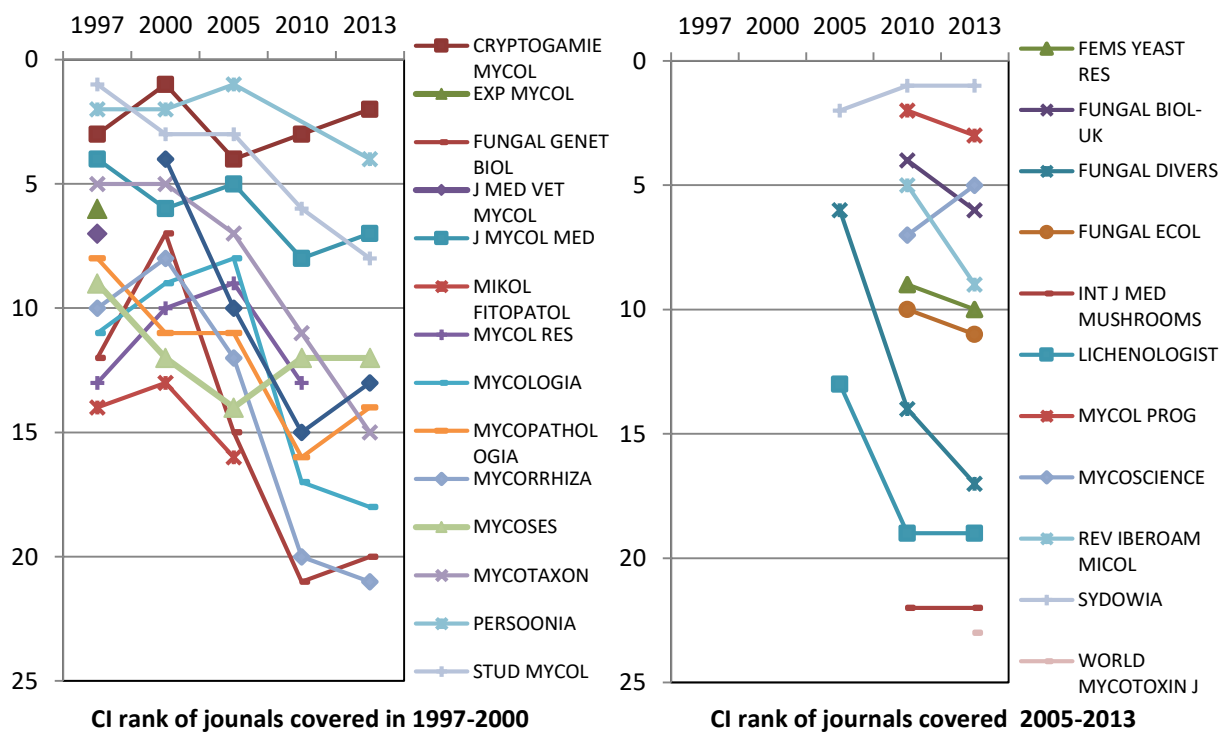

**Figure 3. Relationship between competitive intensity (CI) and time**

Figure 3 shows the difference between the CI rankings for a set of journals between 1997 and 2000, and a second set of journals between 2005 and 2013. For the first set, the CI rankings for most of the 14 journals decreased from 1997 to 2013, and only four were in the top ten. This typically means that the competitive pressures of traditional journals (with a longer publishing history) were declining. At the same time, most of the second set started in a high competitive pressure situation, and approximately half of them remained in the top ten of the CI ranking. This means these new journals had to face more challenges.

*Competitive intensity for a subject*

Equation (6) shows that the CI for a subject is the average competitive pressure of all the journals. We calculated the CIs for mycology in 1997, 2000, 2005, 2010, and 2013.

**Table 3. Competition intensity (CI) and number of journals for mycology**

|  | *1997* | *2000* | *2005* | *2010* | *2013* |
| --- | --- | --- | --- | --- | --- |
| number of journal | 15 | 14 | 17 | 23 | 23 |
| CIS | 29.489 | 39.110 | 43.726 | 38.500 | 41.361 |

Table 3 shows that the competitive intensity for a subject (CIS) increased from 1997 to 2005, but the number of journals only increased from 15 to 17. We can see that the CIS decreased between 2005 and 2010 because the number of journals increased from 17 to 23 (by approximately 35%). By analysing the relationship between the subject's scale and CIS, we can see that more journals correspond to low CIs. From 2010 to 2013, the number of journals was stable at 23 so the CIS increased. In the absence of any other influences, the CIS will continue to increase.

By analysing the competitive pressure on each journal and the CIS, we can determine the state of the competitive environment using a quantitative method, and compare the competitive relationships of different journals and subjects. Through a comparative analysis, we can research reasons for any differences and provide S&T publications with scientific data and tools. Additionally, the data can be used to monitor the S&T journals environment at a macro level, and help decision makers with regard to administration.

**Conclusions**

*There is vast difference in the CIs between subjects and competition pressures between journals.*

We have measured journals' competition pressures and the CIS using quantitative methods. The differences between journals' competitive environments may be caused by many related factors. Different journal attributes are related to competitive pressure. For example, the competitive environment and resources vary among multidisciplinary, ordinary professional, and specialized professional journals. Fundamental research or academic journals and engineering or application journals have different competitive features. Chinese journals are obviously different to English language journals. So the factors that influence competitive pressure and intensity, measurements of these related factors, and mechanisms that influence journals' competitive environments must be studied further.

*The competitive pressure from a powerful rival may be equal to the pressure from several weakly similar journals.*

The ecological concept of CI is a combination of all kinds of competitive pressure. So the competitive pressure on a journal is a combination of the competitive pressure from all of its rivals. The competitive pressure from a powerful rival may be equal to the pressure from several weakly similar journals. The combination of competitive pressure for each journal may be different, which can lead to a high competitive pressure and number of rivals. It can be used as reference when analysing a target journal's competition.

A journal's homogeneity is important when developing S&T journals. Using our quantitative method, we found that homogeneity is obvious in some fields, especially journals that lack "personality". Such journals have higher competitive pressures. The homogeneity of a journal increases its competitive pressure, and the homogeneity of a subject hinders a favourable competitive environment. There is typically fierce competition between two journals that are very similar. Abnormal cooperative relationships exist between some journals, who adopt inter-citation journal group models. These very similar journals pursue high IFs and cited rates. The academic misconduct phenomenon is one problem that results from a journal's homogeneity.

*More study is required for multidisciplinary or interdisciplinary journals.*

In our method, each journal only belongs to one subject. However, developments in science and technology have led to fusions and evolutions in subject areas. Most articles belong to more than one subject area. At the same time, some journals are multidisciplinary, so it can be difficult to define their subject. We measured a journal's competitive pressure in terms of only one subject. Future research is required to determine how to measure and compare competitive pressure and similarities for multidisciplinary or interdisciplinary subjects.

*A favourable competitive environment is only possible at the proper scale*

The scale of the subject (number of journals) is related to its competitive pressure and intensity. A favourable competitive environment is only possible at the proper scale. If there are too many or too few journals the CI decreases. In S&T journal administration, the distribution and trends of the CIs can be used as a reference to promote the development of favourable and sustainable environments.

*Risk of method in application: uncontrollable noise data*

In theory, this method is assuming a group covering all journal in subject. But in the actual evaluation process, will find the subject of inaccurate information. Should not appear there may be journals and lack of important journals in subject because of reasons. In application, which may become uncontrollable noise data, and disturbing results. In many cases, this may be that different researcher and analyst have distinct understanding about the definition of disciplinary items and disciplinary system structure. This paper uses the JCR disciplinary system but it is not always the only best selection for each research case.

*The research findings in this study can be used as a reference for a new journal when choosing a subject and field.*

In management science, there are "red ocean" and "blue ocean" strategies when facing competitive environments. The red ocean strategy directly reacts to competition, whereas the blue ocean strategy avoids direct competition and exploits new markets (Chan and Mauborgne 2005). When facing competition from rivals, S&T journals must choose an optimal path based the current environment and future positioning. Journals with relative advantages tend to use red ocean strategies, proactively consolidating and extending their advantages. Relatively weak journals use blue ocean strategies, seeking paths that reduce homogeneity problems and competitive pressures. The findings of this study can be used as a reference for a new journal when choosing a subject and field. In a fiercely competitive fields, it is difficult to successfully launch a new journal without obvious diversity. Academic journals should pursue most valuable academic papers published. A journal from obscurity to academic publishing giant is fraught with competition. Measure of the competition environment, help the academic journals to find the most efficient development path.

**Acknowledgments**

This research was supported by National Social Science Foundation of China (Project Number:15BTQ059) and National Key Technology Support Program of China (Project Number: 2015BAH25F01).